**Title: 8 quick tips for data-model integration in ecology**

Short title: Data-model integration in ecology

## Authorship

Laurinne J Balstad[1,2,*] (ORCID 0000-0002-4615-529X), Joe Brennan[2,3,*] (ORCID 0009-0006-4178-9919), Marissa L. Baskett[1,2,4] (ORCID 0000-0001-6102-1110), Mattea K. Berglund[5] (ORCID 0000-0002-3076-9465), Mei Z. Blundell[1,2,4] (ORCID 0000-0003-2544-2895), Jessica A. Bolin[4,5] (ORCID 0000-0002-9868-7511), Amy A. Briggs[2,3] (ORCID 0000-0002-5782-7928), Mary C. Fisher[1,4] (ORCID 0000-0003-0729-3929), Christopher M. Heggerud[1,2,6] (ORCID 0000-0002-0497-9506), Madeline Jarvis-Cross (ORCID 0009-0008-0527-9369)[1,4], Lauren Mossman[1,4,7] (ORCID 0009-0008-5700-7170), Andrea N. Odell[1,8] (ORCID 0009-0008-7780-3895), Jennifer Paige[1,4,7] (ORCID 0000-0002-5753-0403), Sophia Pelletier[5] (ORCID 0009-0005-9986-3117), Mikaela M. Provost[5] (ORCID 0000-0001-8301-515X)

**Shared first authorship*

1. Department of Environmental Science and Policy, University of California, Davis USA 95616;
2. Center for Population Biology, University of California, Davis USA 95616;
3. Department of Evolution and Ecology, University of California, Davis USA 95616;
4. Coastal and Marine Sciences Institute, University of California, Davis USA 95616;
5. Department of Wildlife, Fish, and Conservation Biology, University of California, Davis USA 95616;
6. Department of Mathematics, University of Manitoba, Winnipeg Canada R3T 2N2;
7. Department of Mathematics, University of California, Davis USA 95616
8. California Department of Fish and Wildlife, West Sacramento USA 95605

"Data! Data! Data! I can't make bricks without clay." – Arthur Conan Doyle, *The Adventure of the Copper Beeches*

## Introduction

Theoretical ecologists have long leveraged empirical data in various forms to advance ecology (1). Recently increased volumes and access to ecological data present an expanding set of opportunities for theoreticians to inform model development, framing, and interpretation. Whereas statisticians have collective guidance on best practices for data use (e.g., (2)), theoreticians might lack formal education on how to integrate diverse types of data into a single ecological model.

Developing mathematical theory is a parallel process to empirical work (3,4). Modelers start by identifying and narrowing the research question, then building, analyzing, and interpreting their model. The development interpretation of models helps develop further research questions and hypotheses to be tested through multiple means. For dynamical systems modeling, theoreticians translate biological systems into mathematical equations, using state variables to represent quantities that change over time and parameters as predefined, usually fixed quantities that determine how state variables change over time. This translation process requires bringing together multiple types of qualitative and quantitative data to carefully articulate how systems function (Figure 1).

Here, we have developed a guiding framework to support theoretical ecologists in synthesizing multiple types of data at different phases of the modeling process (Figure 1). Throughout, we point to helpful references that provide different possible data-model integration pipelines (e.g., (5,6)). Additionally, we show the application of these tips to two specific cases, guiding readers through how these tips can be used across the modeling process and in different contexts (Box 1). Our tips fall into three overarching themes: iteration in the data-model integration process (Tip 1, 8; (5)), leveraging multiple sources of data (Tips 2-5; (7)), and understanding uncertainty (Tips 6-7; (1,8)). These tips point towards cross-talk between modelers and empirical scientists as a way to improve modeling efforts and advance scientific knowledge. Across these tips, we emphasize that data-model integration requires transparent, justifiable, and defensible communication of modeling choices to help readers appropriately contextualize the model and its implications.

**Tip 1: Make the data-model integration process iterative**

Though workflows for data-model integration can appear linear in academic writing, the process commonly requires revisiting prior steps when new insights are gained from later steps (5). Across the data-model integration workflow, there are multiple places to incorporate different data sources and types, which can help support modeling choices and refine or revise models (Figure 1). For example, if a constructed model produces dynamics that do not align with expected system dynamics, then a theoretician might revisit a second time the scientific literature, model formulation, or parameterization and subsequently adjust their model or re-frame their model interpretation (e.g., in light of possible limitations; see example in Box 1). In model development, drawing on established knowledge from qualitative and quantitative data can help revise the question framing, model structure, and parameterization to better answer the central question being explored. Moving 'backwards' in the workflow is not regression on the research project, but rather a natural and necessary procedure.

**Tip 2: Integrate multiple types of data**

Diverse data types can shape and inform ecological models (Table S1). We advocate for using multiple forms of data throughout the modeling process, recognizing that data goes beyond quantitative values (Figure 1). Mental models and intuition can shape model conceptualization and interpretation (9), while qualitative data can guide model framing and reveal dynamics not captured by quantitative datasets. Quantitative data can inform parameterization and sensitivity analyses. Expert opinion can help ground model formulation, parameterization, and outputs.

We also encourage theoreticians to consider whether and how to bring together multiple ways of knowing (*sensu* (10,11)) in their research (12). Co-production and collaboration with empiricists, social scientists, and knowledge holders outside academia can identify new research questions, reveal new hypotheses, and broaden model relevance (7). For example, MacCall et al. (13) co-produced heuristic models of fish population dynamics, drawing on Traditional Knowledge of Pacific herring migration patterns. Bringing together both Western and Traditional Knowledge of the system allowed MacCall et al. to explore the impacts of an otherwise underappreciated mechanism of follow-the-leader migration behavior on fisheries management outcomes (13). Incorporating diverse types of knowledge throughout the modeling process requires more than extracting data; it calls for ethical engagement with and critical reflection on the systems from which data emerge (7). This includes reflecting on the epistemologies (14), historical and present contexts, uncertainties, and biases embedded in how data is defined, collected, transmitted, and interpreted. Rather than integrating other forms of knowledge into Western norms of scientific inquiry, ethical engagement involves respecting distinct knowledge systems and sovereignty, avoiding extractive practices, and supporting Indigenous leadership in co-producing questions and knowledge (15–17). Engagement with Indigenous title-, rights-, and knowledge-holders will ethically, and sometimes legally, require co-production (see Supplemental Glossary; (16,18)). For a more comprehensive discussion of ethical knowledge co-production, we direct the reader to, among others, (16,18–21). Embracing a broader view of data and learning how to ethically engage with diverse data and multiple ways of knowing (10,20) enables more relevant, inclusive, and impactful ecological modeling (22,23).

**Tip 3: Align the use of data to the research goals**

While our ability to access “big data” makes adding more data to models tempting, consider the amount and precision necessary for the project goals (24,25). One framework for categorizing research questions and associated models is May’s tactical/strategic spectrum (1), which can guide the degree of specificity needed for building a model. Tactical questions require detailed, system-specific information for realistic, precise takeaways, sacrificing generality ((1); Levins’s Triangle, (22); Figure S1). Models answering tactical questions might require carefully identifying parameter values from empirical literature or reducing error between quantitative empirical data and model output in order to identify parameter values. For example, Kaare-Rasmussen et al. (27) constructed a dynamic energy budget model considering the relative role of anemone mutualists and parasites in affecting host growth. To parameterize the model and quantify the environmentally mediated effect of symbionts on host growth, they used a maximum likelihood-based approach to minimize the error between experimental time series and their model outputs. In contrast, strategic questions aim for generalizable takeaways rooted in a realistic case, sacrificing precision (1,26). Models emphasizing ecological patterns might use qualitative data to ground model assumptions, coupled with sensitivity analysis to understand how parameters affect outcomes; determining the precise value of a parameter might not be necessary to achieve the model’s goal. For example, Karatayev and Baskett (28) used quantitative and qualitative data to build a kelp forest-inspired model to test whether alternate stable states remained relevant in the presence of an array of dispersal dynamics; in this case, extreme precision in data informing the model’s parameters was not as important to define the model’s structure, given the focus on qualitative outcomes. Letting model goals guide the need for specific pieces of data can help ensure that the data-model integration process sees the forest through the trees. Einstein’s philosophy of “Make it as simple as possible, but no simpler” guides theoreticians when designing models, and we would suggest, in the context of model building, “Use as much data as you need, but no more.”

## Tip 4: Use quantitative and qualitative data to link mathematical form to biological meaning

Theoretical ecologists translate biological systems into mathematical equations and mathematical results into ecological insights. Generally, the first step in building and explaining a mathematical expression is to identify key biological mechanisms within the model (25). This step can be informed by diverse types of data, like mental models informing understanding of a system's functioning or experimental data quantifying the relationship between system components. Then, the next step is to formalize the biological mechanism into mathematics, often drawing on typical mathematical forms (e.g., Holling functional responses for predator-prey interactions) or past mathematical theory directly. In some cases, it might also be possible to derive the functional form from first principles (e.g., (29)) or empirical insights (e.g., (30)). Being clear about relationships between biological data, assumptions, and mathematical forms grounds models in reality and provides bounds for model interpretation. Importantly, multiple models, each with differing underlying data and assumptions, can contribute multiple perspectives on a single biological question, creating a more robust scientific understanding (3,31).

## Tip 5: Gut-check data with others and question your interpretations

Data interpretations can vary as they reflect an individual's perspective, modeling approach, and understanding of the system (7). Erroneous assumptions about data early in model building can lead to incorrect model mechanisms, misinterpretation or bias in results, unrealistic patterns, and poor predictive power (see example in Box 1). A careful, critical gut-check of both the data and the interpretation of it can help avoid these issues in the data-model integration process. While reviewing literature and creating mental models can help support data interpretation, talking frequently to people with system expertise can ensure thoughtful integration. Just as talking with academics requires thoughtful accreditation of ideas and understanding (e.g., CReDIT, (32)), talking with system experts outside of academia similarly requires ethical processes to avoid harmful extraction of data, ensure consent, properly recognize contributions, and build long-term respectful relationships (10,18,19,33,34). A modeler ("I" below) might ask such experts ("you" below) questions like:

- Which processes do you think are most important in driving process X within this population/system?
- Based on the data, I've done Y with my model. Do you think Y represents the data and your understanding of the system?
- Are the challenges I'm facing in building and parameterizing my model reflective of possible gaps in the empirical literature? Is there expert knowledge that I'm not finding that would support my modeling process?
- Do you predict your results would be similar across populations, or is there a unique aspect of the population you studied that leads to these results?
- How did you measure variable Z? What are the units of variable Z? How confident are you that you captured this value?

**Tip 6: Contextualize model results in light of reducible and irreducible uncertainty**

Uncertainty is inevitable in ecological models; it can be "reducible" if more knowledge helps to minimize the uncertainty, or "irreducible" if the uncertainty is an innate part of the process (35). . For instance, a model parameter such as average clutch size could vary due to observation error (reducible uncertainty) or demographic stochasticity (irreducible uncertainty; (35)). In the former case, theoreticians can explore how changes to the parameter mean (i.e., improved accuracy of the measurement) might alter model results; in the latter case, theoreticians can explore how changes to the parameter variance (i.e., degree of stochasticity) impact model results. Discussing the data sources informing model structures and parameters, and the limitations of those choices, and missing knowledge, can help the reader understand the empirical contexts for which the model applies and inspire future research needs (5,24,31). More broadly, theoreticians can address reducible uncertainty with questions like "For the process of interest, what model components require more resolution, and does such precise understanding or measurement exist?" They can address irreducible uncertainty with questions like "What data sources and associated model components drive uncertainty in model estimates?".

**Tip 7: Formally evaluate model uncertainty and prediction sensitivity**

In addition to acknowledging sources of uncertainty, conducting a formal sensitivity or uncertainty analysis (SA/UA; see Supplemental Glossary) helps to systematically describe uncertainty, clarify interpretation of the model results, and increase model credibility (6,8,25,36,37). It can also inform future research priorities (8,38). SA/UA can be implemented in many ways depending on the model goals, the amount of data, the type of uncertainty of interest, the computational complexity of the model (e.g., number of input factors, interactions, etc.) and the intended application of the model or SA/UA (discussed in (6,8,36,39)). For example, a local, one-at-a-time sensitivity analysis can help describe how changing a set of key input factors, usually established *a priori* based on the central modeling question, influences model outputs. Note, though, that one-at-a-time analysis is limited by its inability to identify non-linear interactions between parameters, which are often central to dynamical systems modeling (40). In contrast, a global sensitivity analysis can identify the effects of interactions between input factors (41), with the potential limitation that any one draw might produce an unrealistic or biologically infeasible system. Another approach might be to test multiple parameterizations (e.g., representative of different possible species or systems) to identify which model results and associated qualitative trends are robust to the various parameter sets.

### Tip 8: Apply the iterative data-model conversation across studies

Data-model integration often extends beyond a single research project. Expanding from Tip 1, we finish our set of tips by recognizing the importance of cross-talk between modeling efforts and data collection across projects. Ideas for new models or clarity about previously developed models can be informed by insights from new data, drawing on all the dimensions of data we have highlighted here (Tips 2-4). Inversely, theoretical ecology can help inform future data collection through modeling exploration (Tip 6) and UA/SA (Tip 7), which can identify the mechanisms and parameters that have outsized influence on results, indicating the need for more precise estimation. Cross-talk between theory and data collection (Tip 5, Figure 1) is necessary to inform our understanding and interpretation of results derived from both theoretical and empirical work: "A model without data is empty; data without a model is blind" ((42), pg. 72).

## Conclusion

Some uncertainty or imprecision is inevitable in both the structure and parameterization of theoretical models. Working with data that are biased, uncertain, imprecise, or incomplete does not render a model useless. As demonstrated by our own experiences from past modeling projects (described in Box 1), iterating across the data-modeling integration process (5), leveraging diverse data sources (7), and understanding uncertainty (1,8) can help theoreticians situate their models within the best available scientific context, support appropriate model interpretation, and identify future directions for empirical and theoretical research.

## Elements

*Figure 1. A flow diagram illustrating how data can inform the modeling process and vice versa.*

The leftmost column (A) broadly identifies two data types, quantitative (yellow-orange) and qualitative (pink). The rightmost column (C) outlines the iterative modeling process, with dark black arrows demonstrating the way a theoretician might move through or revisit modeling steps. The central column (B) illustrates how the data-use process joins data to the modeling process. From (A), certain data types naturally lend themselves to particular uses; we use the box color to reinforce the use of different types of data for different purposes, where the gradient-colored boxes indicate both data types might have common use. Flowing to (C), the uses of data from (B) allow progression of the modeling process by providing key pieces of information. This flow diagram represents one, non-exhaustive way a theoretician might leverage data to move through the modeling process. Tips associated with different figure components are labeled in parentheses with “T” and the Tip number.

*Box 1. Putting the tips to action.* Here, we provide the behind-the-scenes process of building, parameterizing, and interpreting theoretical models from our own experiences to highlight how our tips can be used by theoreticians to create more informed–and more informative–models.

*Tips 1, 2, 5 and 6:* In Serpico et al (43), the authors (including CMH) built a model for aquatic population dynamics to describe the effects of lake eutrophication and warming scenarios at different trophic levels. The model was parameterized using time series data of dissolved oxygen concentration, population dynamics, and various other lake characteristics. While the data suggested that organisms could survive extremely low dissolved oxygen concentrations, the authors’ own literature scan and mental models seemed to contradict this information (Tip 2). The authors thought maybe the time series were inaccurate. However, through further literature search and discussion with others (Tips 1, 6), the authors realized they were modeling a “well-mixed” environment, while the data was collected from the bottom of the lake, where water tends to be stagnant and have less dissolved oxygen. To support readers in understanding the uncertainty and limitations of the model (Tip 6), the authors pointed out the key ecological contexts to which their model would and would not apply (i.e., lake systems without and with spatial structure, respectively).

*Tips 1, 3, 6, 7, and 8:* In Baskett et al (44), the authors (including MLB) investigated how including (*versus* excluding) evolutionary dynamics for coral symbionts affected anticipated coral reef responses to future climate change. While data existed for the coral demographic and ecological parameters, no data were available to independently parameterize the strength of selection on the thermal tolerance of coral symbionts. Therefore, the authors tuned this parameter to qualitatively match observations of when coral declines occurred given local temperature trajectories for select locations. Because of this qualitative approach with uncertainty in exact parameter values, the authors framed their interpretation around qualitative insights rather than emphasizing the model output as a precise prediction (Tip 3). When examining the output model dynamics, they observed seasonal fluctuations in symbiont dynamics, an outcome that was unplanned. Following this observation, the authors revisited the literature to see if this was realistic and found empirical evidence for such dynamics, which provided an unexpected opportunity for model validation (Tip 1). A local sensitivity (elasticity) analysis ranked the parameters' relative influence on coral cover, which helps guide priorities for empirical research that might enable more quantitatively precise predictions (Tips 7-8). Finally, an exploration of multiple climate models and climate scenarios, which capture scientific and societal uncertainty in climate projections (Tips 6-7), elucidated the central takeaway of the paper: evolution can promote coral persistence under more moderate, but not more severe, climate scenarios.

## CRediT Statement

LJB contributed to conceptualization. LJB, JB, MLB, and MMP contributed to investigation and project administration. MLB and MMP contributed to supervision. LJB, MLB, MZB, MMP contributed to visualization. LJB, JB, MLB, MKB, MZB, JAB, AAB, MCF, CMH, LM, ANO, JP, SP, and MMP contributed to writing - original draft. All authors contributed to writing - review and editing.

## Competing interests

The authors have no competing interests to declare.

## Funding information

LJB acknowledges support of National Science Foundation National Research Traineeship award #1734999 and National Science Foundation Graduate Research Fellowship Program award #2036201. JB acknowledges support of National Science Foundation Graduate Research Fellowship Program award #2439024. MZB acknowledges support from National Science Foundation National Research Traineeship award #1734999 and National Science Foundation Graduate Research Fellowship Program award #2439024. CMH acknowledges support from Natural Sciences and Engineering Research Council of Canada, Postdoctoral Fellowship. MJC acknowledges support from National Science Foundation Grant #2025235. ANO acknowledges support of NMFS-Sea Grant Population and Ecosystem Dynamics Fellowship #NA23OAR4170535 and National Science Foundation National Research Traineeship #1734999. JP acknowledges the support of the U.S. Department of Energy, Office of Science, Office of Advanced Scientific Computing Research, Department of Energy Computational Science Graduate Fellowship under Award Number DE-SC0024386. The funders had no role in study design, data collection and analysis, decision to publish, or preparation of the manuscript.

**Acknowledgments**

We would like to acknowledge Jacques Costeau Baskett for support throughout the writing process. Thank you to Maya Chari, Zoe Douglas, and Hunter Milles, and three anonymous reviewers for helpful comments that improved the manuscript.

*Text S1.* Supplemental Glossary: A list of terms and corresponding definitions used in the manuscript

*Figure S1.* Description of the precision-generality-realism triangle

*Table S1*. A non-exhaustive sample of how various data can be used in ecological modeling.